\documentclass[conference, 10pt]{IEEEtran}
\usepackage{graphicx} 
\usepackage{xcolor,soul,framed}
\usepackage{amsfonts}

\colorlet{shadecolor}{yellow}
\usepackage{kotex}
\usepackage{amsmath}
\usepackage{array}
\usepackage{mdwmath}
\usepackage{mdwtab}
\usepackage{eqparbox}
\usepackage{url}
\usepackage{bbm}
\usepackage{amssymb}
\usepackage{tikz}
\usepackage{subfigure}
\usepackage{cite}

\usepackage{titlesec}

\DeclareRobustCommand\encircle[1]{\tikz[baseline=(char.base)]{\node[shape=circle,fill,inner sep=1pt] (char) {\textcolor{white}{#1}}}}

\begin{document}
\title{\LARGE Deadline-Aware Bandwidth Allocation for \\ Semantic Generative Communication with Diffusion Models}
\author{\IEEEauthorblockN{Jinhyuk Choi$^{1}$, Jihong Park$^{2}$, Seungeun Oh$^{1}$, and Seong-Lyun Kim$^{1}$}\\

\IEEEauthorblockA{$^1$School of EEE, {Yonsei University}, Seoul, Korea, email: \{jh.choi, seoh, slkim\}@ramo.yonsei.ac.kr}
\IEEEauthorblockA{$^2$SUTD, {Singapore University of Technology and Design}, Singapore, email: jihong\_park@sutd.edu.sg}
}

\maketitle

\begin{abstract}
{T}{he} importance of Radio Access Network (RAN) in support Artificial Intelligence (AI) application services has grown significantly, underscoring the need for an integrated approach that considers not only network efficiency but also AI performance. In this paper we focus on a semantic generative communication (SGC) framework for image inpainting application. Specifically, the transmitter sends semantic information, i.e., semantic masks and textual descriptions, while the receiver utilizes a conditional diffusion model on a base image, using them as conditioning data to produce the intended image. In this framework, we propose a bandwidth allocation scheme designed to maximize bandwidth efficiency while ensuring generation performance. This approach is based on our finding of a ``\emph{Semantic Deadline}''--the minimum time that conditioning data is required to be injected to meet a given performance threshold--within the multi-modal SGC framework. Given this observation, the proposed scheme allocates limited bandwidth so that each semantic information can be transmitted within the corresponding semantic deadline. Experimental results corroborate that the proposed bandwidth allocation scheme achieves higher generation performance in terms of PSNR for a given bandwidth compared to traditional schemes that do not account for semantic deadlines.
\end{abstract}

\begin{IEEEkeywords}
Generative AI, diffusion model, image inpainting, semantic communication, bandwidth allocation. 
\end{IEEEkeywords}

\section{Introduction}
\IEEEPARstart{T}{he} emergence of Artificial Intelligence Generated Content (AIGC) services has gained significant attention in recent years, driven by their transformative potential across diverse industries. These services, which leverage generative AI to generate text, images, videos, and audio, offer unprecedented efficiency and creativity in content creation. These advances in AIGC have led to the attention of semantic generative communication (SGC) frameworks to support this application in wireless networks, where the transmitter transmits semantic information and the receiver directly or indirectly utilizes it to generate the intended image through generative AI \cite{tong2022nine ,grassucci2024generative, choi2023enabling, xu2024unleashing}. Despite this intensive interest, existing research lacks a communication system design that accounts for both network efficiency and the unique characteristics of AIGC services. However, this requires a comprehensible understanding of the correlation between semantic information and intended image in generative AI, which makes it non-trivial, especially when dealing with multi-modal semantic information.

To explore a multimodal AIGC framework, our research focuses on designing uplinks over SGC, specifically for image inpainting applications using a conditional diffusion model. In this framework, the transmitter’s objective is to deliver the intended image to the receiver—not by directly transmitting it, as in traditional communication, but by sending semantic information as in semantic communication. This information, derived from a semantic mask and textual description, conveys the transmitter’s intent—specifically, how the base image (shared between transmitter and receiver) should be modified. The receiver then generates the intended image by applying the diffusion model, using the base image as input and the semantic information as conditioning data.

In the multimodal SGC framework described above, we empirically find that each piece of semantic information has a specific ``\emph{Semantic Deadline}''. Delivering semantic information to the receiver within this deadline maximizes bandwidth efficiency while ensuring the desired generation performance. Building on this finding, we propose a bandwidth allocation scheme for the uplink that accounts for both AIGC performance and communication efficiency. Through numerical evaluation, we demonstrate that the proposed scheme achieves the highest PSNR at a given bandwidth.

\begin{figure*}[t!]
\centering
\includegraphics[width=18cm]{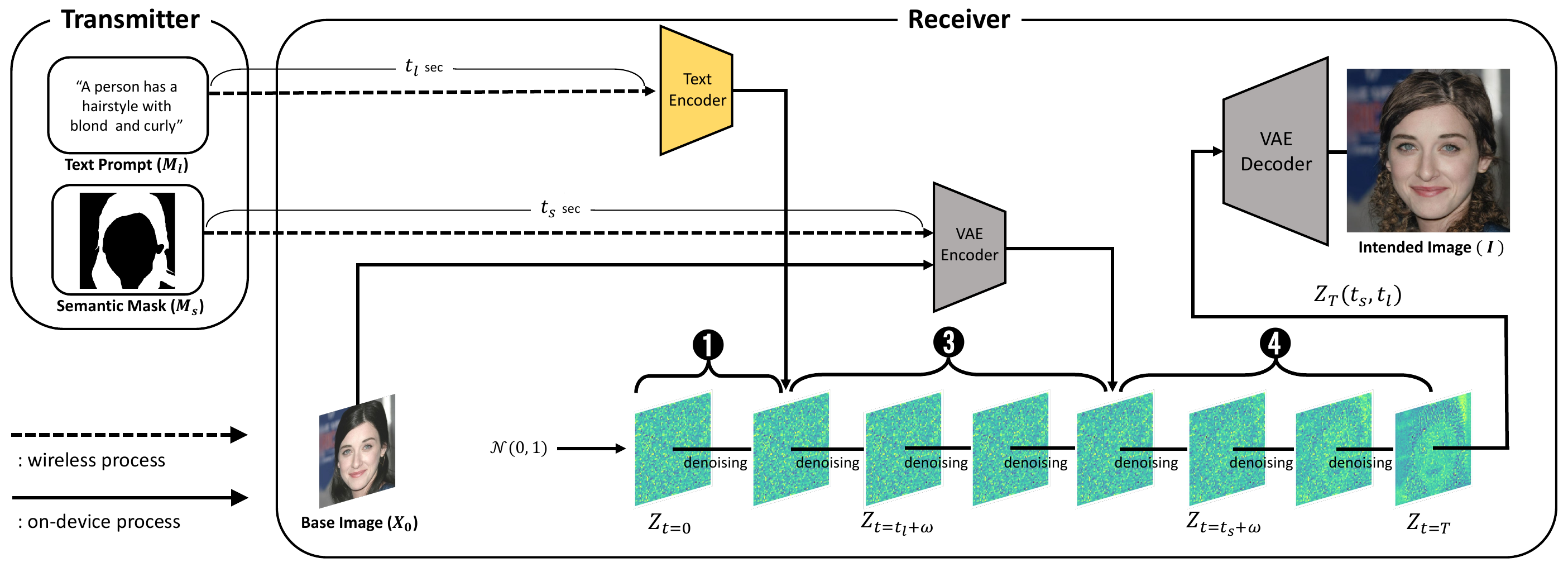}
\caption{System diagram illustrating the overall process. The transmitter sends each semantic information, while the receiver’s internal image inpainting process is depicted, highlighting the asynchronous reception.}
\label{fig:1}
\end{figure*}

\subsection{Related Works and Motivations}
Recent studies have increasingly focused on the multi-modal data importance. In \cite{park2018urllc}, the authors propose a network design methodology that enhances the performance of Virtual Reality (VR) services by considering the relationship between various modalities—specifically visual and haptic—and overall service performance. More recent work, such as \cite{cicchetti2024language}, explores AI/ML-based services by decomposing image transmission into two distinct modalities and analyzing the final output based on whether each modality is transmitted. Furthermore, \cite{qiao2024latency} introduces an adaptive modulation and coding scheme tailored to the channel conditions experienced by each modality during transmission to produce images that closely resemble the original. Although many studies have investigated wireless communication performance in multi-modality transmissions, most treat the computational processes within these services as separate, independent components. In contrast, our work addresses the intersection of transmission and computation domains by leveraging the structural characteristics of diffusion model. The proposed design demonstrates strong potential for practical deployment in real-time multimodal generation scenarios under bandwidth-limited wireless systems.

\subsection{Contributions}
The contributions of this work are summarized as follows.
\begin{itemize}
    \item We define a novel metric, \textbf{Semantic Deadline}, to quantify the timeliness requirement of each semantic element for generating the final output.
    \item Leveraging this concept, we design a \textbf{heuristic uplink resource allocation algorithm}, outperforming baselines that ignore semantic significance.
    \item Our proposed design improves resource efficiency, with simulations validating its potential for future wireless systems.
\end{itemize}

\section{System Model and Problem Description}
In this section, we present the semantic generative communication (SGC) framework for multi-modal image inpainting. A shared base image is modified using semantic inputs—namely, a mask and a text prompt. We aim to optimize bandwidth allocation under resource constraints to maximize inpainting quality, enabling real-time, semantic-aware transmission. The overall structure is illustrated in Fig.~\ref{fig:1}.

\subsection{System Model}
In the proposed SGC framework, a single transmitter and receiver communicate over two independent channels. The receiver runs a Stable Diffusion-based image inpainting model \cite{rombach2022high}, where semantic inputs—mask and text—arrive asynchronously due to varying channel conditions. This asynchrony leads to out-of-sync conditioning during the diffusion process, causing the inpainting result to adapt dynamically to the reception order and delay. Further details are provided in the following sections.

\subsection{Transmitter}
The transmitter sends two types of semantic information: a binary semantic mask \(\boldsymbol{M}_s\) and a textual description \(\boldsymbol{M}_l\). The mask \(\boldsymbol{M}_s\) highlights regions to be modified in white and preserved regions in black, and is source-coded to reduce its size before transmission. The text \(\boldsymbol{M}_l\) describes the inpainting target and is encoded into an 8-bit binary sequence using ASCII. The corresponding data sizes are denoted by \(D_s\) and \(D_l\), respectively.

\subsection{Transmission Model}
The semantic mask \(\boldsymbol{M}_s\) and textual description \(\boldsymbol{M}_l\) are transmitted over two orthogonal wireless channels to prevent interference. Each channel undergoes Rayleigh block fading. The received signal for each modality is modeled as
\begin{align}
    \tilde{\boldsymbol{M}}_i = h_i \boldsymbol{M}_i + \boldsymbol{n}_i, \quad i \in \{s, l\}
\end{align}
where \(h_i \in \mathbb{C}\) denotes the complex channel gain, and \(\boldsymbol{n}_i \sim \mathcal{CN}(0, \sigma^2)\) is the additive white Gaussian noise (AWGN). Channel equalization is assumed to compensate for fading, yielding \(\hat{\boldsymbol{M}}_i \approx \boldsymbol{M}_i\).
Despite minimal signal distortion, differences in channel conditions induce asynchronous reception of \(\boldsymbol{M}_s\) and \(\boldsymbol{M}_l\). The reception time for each modality is given by
\begin{align}\label{transmission_delay}
    t_i = \frac{D_i}{B_i \log_2(1 + \gamma_i)}, \quad i \in \{s, l\}
\end{align}
where \(\gamma_i = \frac{|h_i|^2 P}{N_0}\) is the signal-to-noise ratio (SNR), \(P\) is the transmit power, and \(N_0\) is the noise power spectral density. Bandwidth \(B_i\) is allocated independently for each modality. The transmitter is assumed to have sufficient power to adapt \(P\) based on channel quality.

\subsection{Receiver}
The receiver runs a Stable Diffusion-based inpainting model. Upon connection, it initiates reverse diffusion starting from latent variable \(\boldsymbol{Z}_{t=0}\), sampled from \(\mathcal{N}(\sqrt{\bar{\alpha}_k} \boldsymbol{X}_0, (1-\bar{\alpha}_k)\boldsymbol{I})\), where \(\boldsymbol{X}_0\) is the VAE-encoded base image and \(\bar{\alpha}_k\) is a fixed noise scaling factor.

Due to asynchronous reception of the semantic mask and text description, the reverse diffusion process follows four stages:

\begin{enumerate}
     \item[\encircle{\footnotesize 1}] $t < t_s$ \& $t < t_l$ : This is the case when neither the semantic mask nor the textual description has arrived at the receiving end.
    \begin{align}
        \boldsymbol{Z}_{t+\omega} = p_\theta(\boldsymbol{Z}_t, \boldsymbol{M}_s^{rand}, \boldsymbol{M}_l^{rand}).\nonumber
    \end{align}

    \item[\encircle{\footnotesize 2}] $t \geq t_s$ \& $t < t_l$ : In this case, the semantic mask is received, but the textual description is not.
    \begin{align}
        \boldsymbol{Z}_{t+\omega} = p_\theta(\boldsymbol{Z}_t, \boldsymbol{\hat{M}}_s, \boldsymbol{M}_l^{rand}).\nonumber
    \end{align}

    \item[\encircle{\footnotesize 3}] $t < t_s$ \& $t \geq t_l$ : In this case, the textual description is received, but the semantic mask is not.
    \begin{align}
        \boldsymbol{Z}_{t+\omega} = p_\theta(\boldsymbol{Z}_t, \boldsymbol{M}_s^{rand}, \boldsymbol{\hat{M}}_l).\nonumber
    \end{align}

    \item[\encircle{\footnotesize 4}] $t \geq t_s$ \& $t \geq t_l$ : In this case, all data is successfully received.
    \begin{align}
        \boldsymbol{Z}_{t+\omega} = p_\theta(\boldsymbol{Z}_t, \boldsymbol{\hat{M}}_s, \boldsymbol{\hat{M}}_l).\nonumber
    \end{align}
\end{enumerate}

The function \(p_\theta\) denotes the U-Net in Stable Diffusion with weights \(\theta\), and \(\omega\) is the step duration for reverse diffusion. Placeholder inputs \(\boldsymbol{M}_s^{\text{rand}}\) and \(\boldsymbol{M}_l^{\text{rand}}\) are used until the true semantics arrive. The process iterates over a fixed time horizon \(T\), yielding \(\boldsymbol{Z}_T(t_s, t_l)\), which reflects the asynchronous arrivals of \(t_s\) and \(t_l\). Finally, \(\boldsymbol{Z}_T(t_s, t_l)\) is decoded by the VAE to produce the output image \(\boldsymbol{I}\).

\begin{figure}[t!]
    \centering
    \includegraphics[width=8.8cm]{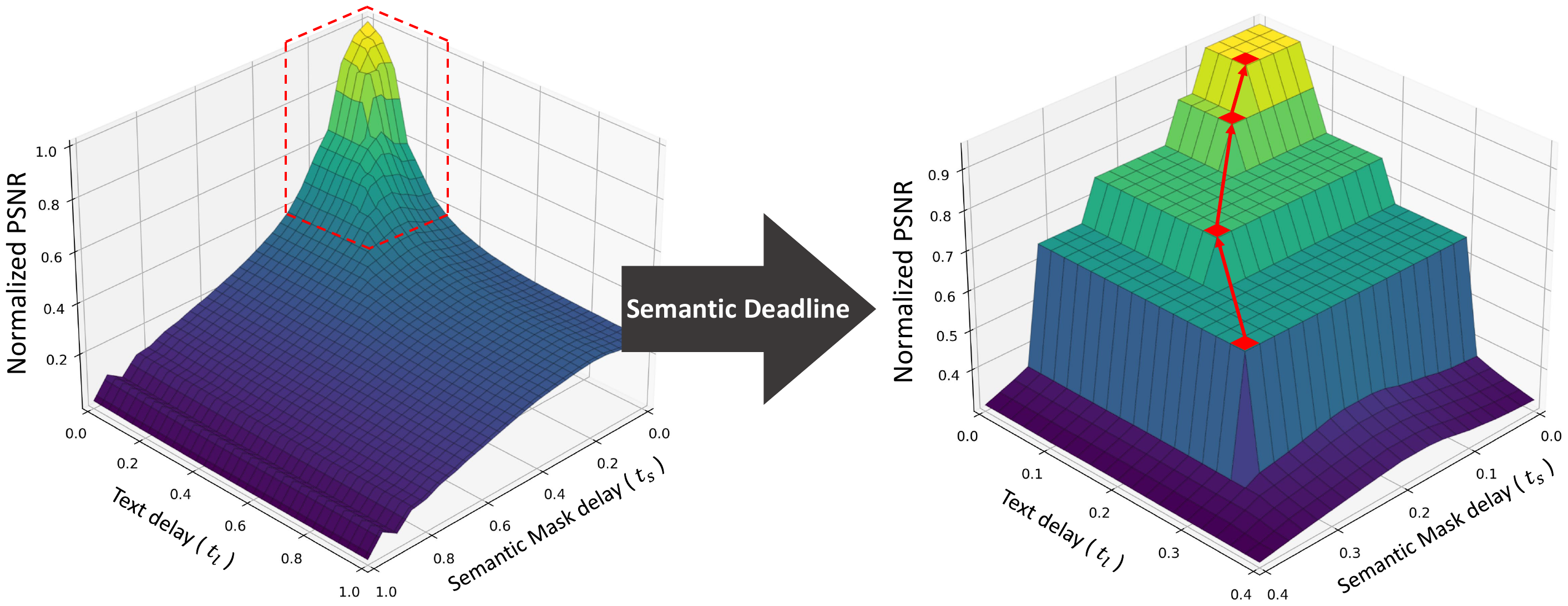}
     \caption{The left graph shows the average $\text{PSNR}(t_s, t_l)$ result. The right graph illustrates the results of semantic deadline clipping with threshold $\varepsilon = \{0.95, 0.85, 0.75, 0.65\}$, aimed at inferring the relationship between the modalities in $\text{PSNR}(t_s, t_l)$.}\label{PSNR_clipping}
\end{figure}

\subsection{Performance Metric}
To evaluate the impact of asynchronous semantic arrivals, we compare the final latent state \(\boldsymbol{Z}_T(t_s, t_l)\) with the ideal reference \(\boldsymbol{Z}_T^* = \boldsymbol{Z}_T(0, 0)\), where both modalities arrive without delay. The quality degradation is quantified using the Peak Signal-to-Noise Ratio (PSNR):
\begin{align}
    \text{PSNR}(t_s, t_l) = 10 \cdot \log_{10}\left(\frac{\text{MAX}^2}{\lVert\boldsymbol{Z}_T^*-\boldsymbol{Z}_T(t_s, t_l)\rVert_2}\right) \nonumber
\end{align}
where \(\text{MAX} = 6\), assuming latent values lie within \([-6, 6]\).

\subsection{Problem Description}
In the proposed system, image inpainting quality improves as \( t_s \) and \( t_l \) decrease, with optimal performance achieved when both approach zero. Since transmission time inversely depends on allocated bandwidth, efficient resource allocation becomes critical under limited wireless capacity. We thus formulate the following optimization problem:
\begin{align}
    &\hspace{-3cm} \textbf{P1:} \max_{\{B_s, B_l\}} \quad \text{PSNR}(t_s, t_l)\\
    &\hspace{-3cm} \text{s.t.} \quad \sum_{i \in \{s, l\}} B_i \leq B \tag{3a}\\
    &\hspace{-1.5cm} B_i \geq 0,\quad i \in \{s, l\} \tag{3b} \nonumber
\end{align}
Here, the objective is to maximize reconstruction quality within a total bandwidth constraint \(B\), ensuring non-negative allocation to each modality.

\begin{figure}[t!]
    \centering
    \includegraphics[width=8.8cm]{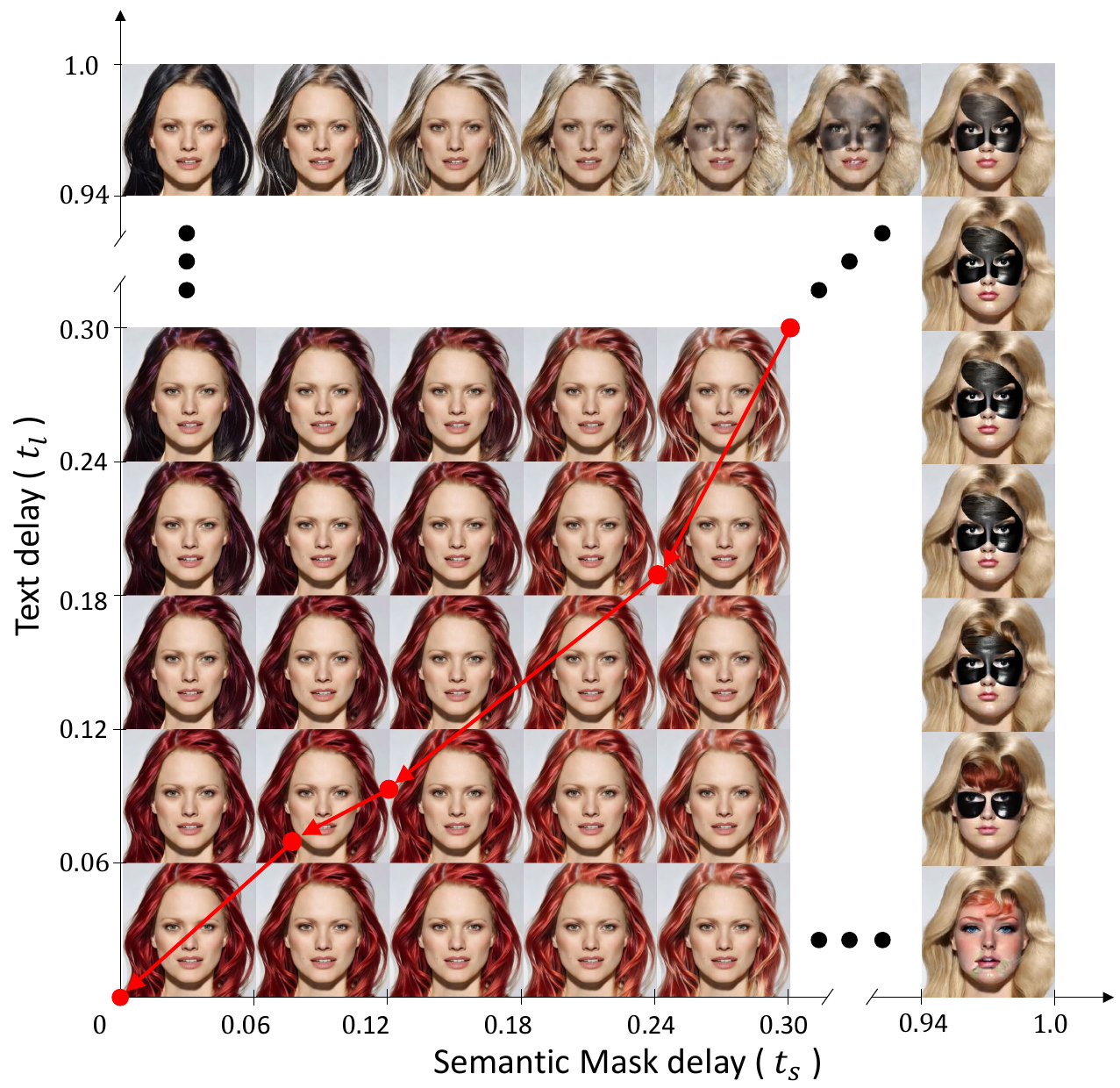}
     \caption{Semantic deadline-aware optimization approach with an example image. Each point represents the initial position set by the semantic deadline point at different thresholds. Semantic deadline indicate the optimization trajectory, illustrating the progression toward improved image inpainting performance.}
\end{figure}

\section{Proposed Solution}
Directly optimizing Problem \textbf{P1} is intractable due to the complex dependencies of inpainting quality on sample-specific attributes, semantic mask coverage, and text content. To overcome this, we reformulate the objective by introducing a performance-bound structure.

\subsection{Semantic Deadline in Image Inpainting}\label{Semantic Deadline}
Through extensive empirical analysis, we observe that $\text{PSNR}(t_s, t_l)$ exhibits a concave, monotonically decreasing trend over the range of perceptually acceptable quality $(\varepsilon \geq \varepsilon_{\text{th}})$. This pattern holds consistently across various image samples and modalities in Stable Diffusion-based inpainting models, suggesting general applicability of the observed behavior.

Let $\mathcal{G}_{\varepsilon}$ denote the set of $(t_s, t_l)$ pairs achieving at least quality $\varepsilon$. These sets are nested:
\begin{align}
    \mathcal{G}_{\varepsilon} \subseteq \mathcal{G}_{\dot{\varepsilon}} \quad \text{if} \quad \varepsilon > \dot{\varepsilon}. \nonumber
\end{align}

Within $\mathcal{G}_{\varepsilon}$, we define the \emph{semantic deadline point} as the farthest point from the origin:
\begin{align}
    (t_{s, \varepsilon}, t_{l, \varepsilon}) = \arg\max_{(t_s, t_l)\in\mathcal{G}_{\varepsilon}}\lVert(t_s, t_l)\rVert_2.
\end{align}

By tracing these deadline points over varying thresholds $\varepsilon \in [\varepsilon_{\text{th}}, 1]$, we construct the set
\begin{align}
    \mathcal{G}^{\star} := \{ (t_{s, \varepsilon}, t_{l, \varepsilon}) \},
\end{align}
which forms a performance boundary curve referred to as the \emph{Semantic Deadline}. This construct provides a model-agnostic basis for relaxing Problem \textbf{P1} into a tractable bandwidth allocation problem. A visual example is illustrated in Fig.~\ref{PSNR_clipping}.

\subsection{Problem Transformation}
Leveraging the semantic deadline, we decomposed \textbf{P1} into two subproblems. First, we aim to determine the maximum achievable threshold $\varepsilon^*$ under the given bandwidth constraints by solving \textbf{P2}. In addressing \textbf{P2}, which is inherently a discrete optimization using (4), we note that the bandwidth constraint does not necessarily satisfy equality in its feasible solutions. Thus, we solve subproblem \textbf{P3} to resolve the optimal allocation values $B_s^*$ and $B_l^*$, ensuring that the bandwidth constraint is met as an equality condition.
\begin{align}\label{P2}
    &\hspace{-3.5cm} \textbf{P2} : \max_{\varepsilon} \quad \text{PSNR}(t_{s, \varepsilon}, t_{l, \varepsilon})\\
    &\hspace{-2.5cm} \text{s.t.} \quad (\text{3a}), (\text{3b}).\nonumber
\end{align}

To find the $\varepsilon^*$, it is necessary to solve the equation (4) for every $\varepsilon$ within the range $[\varepsilon_{\text{th}}, 1]$. If $\varepsilon$ were treated as a continuous variable over this range, the computational complexity would infinite. To mitiage this, we discretize $[\varepsilon_{\text{th}}, 1]$ into $K$ discrete values, reducing the complexity to $O(T^2K)$, where $T$ represents the total denoise steps of the diffusion model.
 
As $\varepsilon^*$ is determined by solving \textbf{P2}, we construct \textbf{P3} as below:
\begin{align}\label{P3}
    &\hspace{-1.3cm} \textbf{P3} : \max_{\{\hat{B}_s, \hat{B}_l\}} \quad \text{PSNR}(t_{s, \varepsilon^*} + \hat{t}_s, t_{l, \varepsilon^*} + \hat{t}_l)\\
    &\hspace{-0.3cm} \text{s.t.} \quad (\text{3a}), (\text{3b}).\nonumber
\end{align}
where $\hat{B}_i$ represents the additional wireless resource allocated to semantic information $i$ beyond the determined bandwidth according to $\varepsilon^*$.

\subsection{Semantic Deadline-Aware Resource Allocation}
In this section, we derive the optimal resource allocations $B_s^*$ and $B_l^*$ by solving \textbf{P2} and \textbf{P3}. First, the optimal solution to \textbf{P2} allow us to determine the minimum image inpainting quality threshold, $\varepsilon^*$, under the given constraints. Consequently, the minimum bandwidth allocated to each semantic information is $B_{i, \varepsilon^*} = \frac{D_i}{t_{i, \varepsilon^*} \log_2(1 + \gamma_i)}, i \in \{s, l\}$. To solve \textbf{P2}, we use the set of $K$ discrete semantic deadline points, denoted as $\mathcal{G}^{\star}(K)$, which is obtained with a computational complexity of $O(T^2K)$. After obtaining $\mathcal{G}^{\star}(K)$, $\varepsilon^*$ can be obtained with an additional complexity of $O(K)$. As $K$ increases, \textbf{P2} provides \textbf{P3} with a more refined optimization scheme, as detailed in the following. 

To solve \textbf{P3}, we employed a heuristic approach guided by the obtained semantic deadlines $\mathcal{G}^{\star}(K)$ in \textbf{P2}. This approach optimizes the allocation of the remaining bandwidth by moving from the minimal resource allocation point toward the next threshold semantic deadline point along the shortest path. By using semantic deadline $\mathcal{G}^{\star}(K)$ that solved by \textbf{P2} and assumption in section \ref{Semantic Deadline}, this approach ensures efficient utilization of the additional resources. The derived optimal bandwidth allocation of each semantic information $B_i^*$ is given by
\begin{align}
    B_i^* = B_{i,\varepsilon^*} + B'\left(\frac{\alpha_{i, \varepsilon^*}}{\sum_{i}^{\{s, l\}}\alpha_{i, \varepsilon^*}}\right), \forall i \in \{s, l\}
\end{align}
where $B_{i,\varepsilon^*}$ represents the minimum allocated bandwidth resource obtained by \textbf{P2}. And $B_{i,grad}$ is given by $\alpha_{i, \varepsilon^*} = \left(\frac{1}{t_{i,\varepsilon^{\prime}}} - \frac{1}{t_{i, \varepsilon^*}}\right)\frac{D_i}{\log_2(1+\gamma_i)}$. Where $\varepsilon^{\prime}$ means the next threshold value compared to $\varepsilon^*$.

\section{Experimental Evaluation}
\subsection{Environment Settings}
We used a Stable Diffusion-based image inpainting model with $T=20$ reverse steps and a guidance scale of $12$. Each step required $\omega = 0.05$ seconds on an RTX 3090 GPU. Experiments were conducted on the \emph{CelebA-HQ} dataset \cite{karras2017progressive}, comprising facial images $\boldsymbol{X}_0$, semantic masks $\boldsymbol{M}_s$ (e.g., hair, eyes, mouth), and corresponding textual descriptions $\boldsymbol{M}_l$. The average data sizes were $D_s = 4$KB (PNG) and $D_l = 1$KB (8-bit ASCII).Wireless channel gains were modeled as i.i.d. Gamma-distributed random variables with unit mean and scale factor $\theta = 2$, i.e., $|h_i|^2 \sim \text{Gamma}(k=0.5, \theta=2)$.

\begin{figure}[t!]
    \centering
    \includegraphics[width=8.8cm]{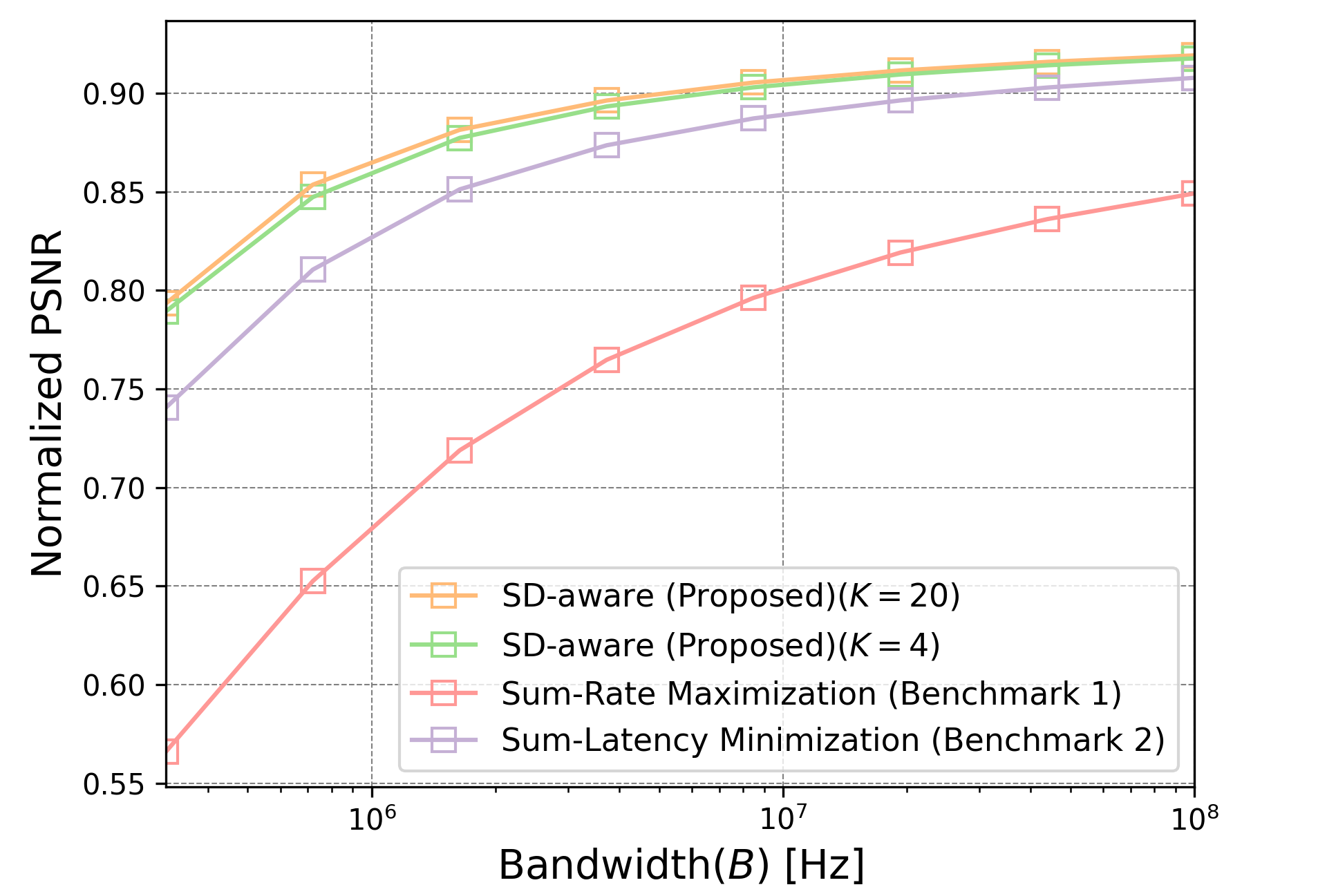}
     \caption{Image inpainting performance under given bandwidth constraints. The figure presents a comparison across four cases: two benchmarks and the proposed algorithm with quantization parameters $K = \{4, 20\}$.}\label{PSNR_bandwidth}
\end{figure}

\begin{figure*}[t!]
\centering
\includegraphics[width=18cm]{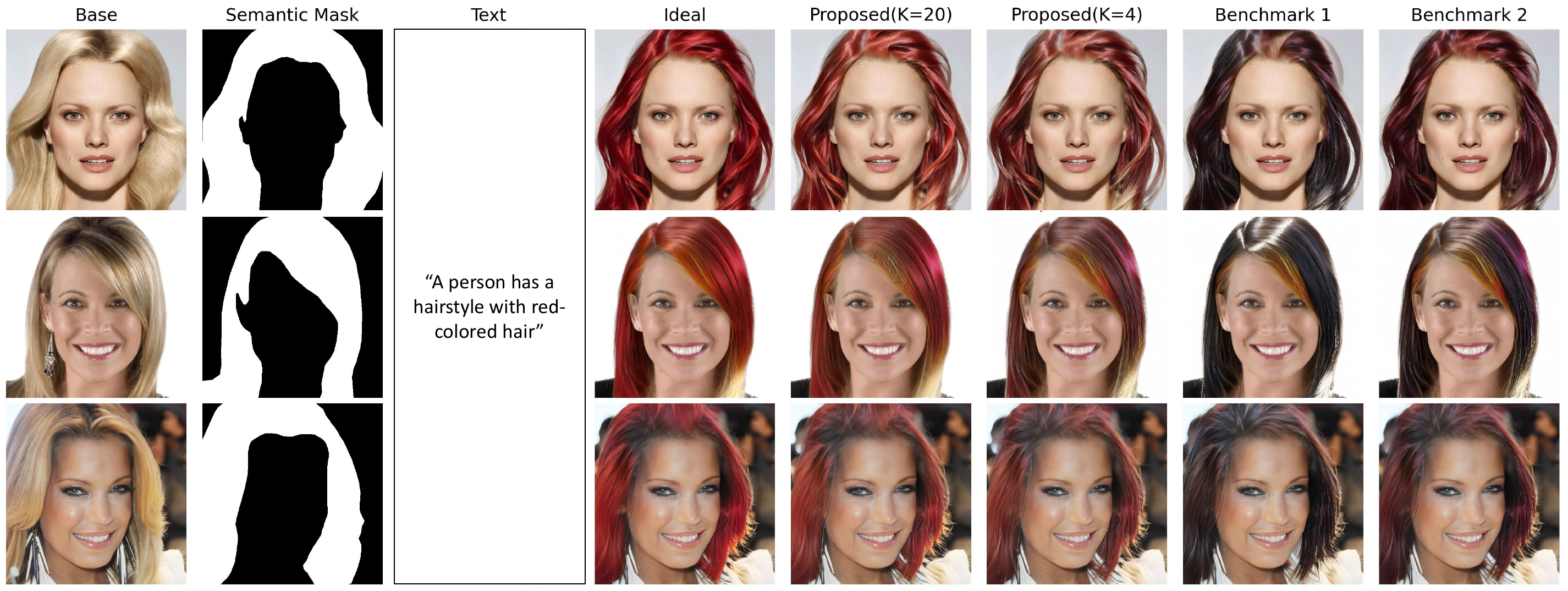}
\caption{Visual comparison of image inpainting results obtained using the proposed algorithm. The ``Ideal'' case represents an optimal result without considering wireless environments, with closer resemblance indicating better performance. In most cases, the proposed algorithm with a quantization parameter of $K=20$ closely approximates the ideal outcome.}\label{visual result}
\end{figure*}

\subsection{Performance Gains under Equivalent Bandwidth}
To verify the effectiveness of our proposed algorithm, we established two benchmark for comparison. Benchmark $1$ aims to maximize the overall throughput across the independent channels. Moreover, benchmark $2$ allocates the available bandwidth resources to minimize the total transmission time of each semantic information $i$ within the given bandwidth constraints. Benchmark $2$, in addition to considering the transmission channel for each semantic information $i$ as in Benchmark $1$, also takes into account the data size. For evaluation, we tested 4 configurations: the two benchmark algorithms and our proposed algorithm with discrete values $K=4$ and $K=20$. As shown in Fig. \ref{PSNR_bandwidth}, our algorithm with $K=20$ guarantee the highest performance for overall bandwidth condition. Benchmark $1$, which optimizes solely based on channel state without regard to semantic information characteristics, yielded the poorest performance. This result underscores the importance of accounting for data attributes in recent semantic communication network. As the discrete value $K$ increases, the performance of the proposed algorithm gradually improves and then converges, as the precise optimization trajectory—modeled by the continuous semantic deadline—reaches a stable direction for bandwidth resource optimization.

\subsection{Visual Quality of the Proposed Algorithm}
To further validate the effectiveness of our proposed algorithm, we visualized its outcomes in Fig. \ref{visual result} to illustrate the impact of its quantitative advantages on actual image results. The visualization is in case with a bandwidth resource $B = 0.3\text{MHz}$, where the transmission channel states for each semantic information are set at $\gamma_s = 2.3\text{dB}$ and $\gamma_l = 3.5\text{dB}$. The visual results reveal distinctions based on discrete value $K$, which previously showed only marginal differences in numerical analysis. For benchmark $1$ and $2$, textual descriptions were not transmitted within the text semantic deadline, resulting in placeholder's wrong image inpainting quality. These visual results highlight the importance of our semantic deadline-aware bandwidth allocation algorithm, demonstrating its advantage in preserving image inpainting service quality. 

\section{Conclusion}
This paper proposed a bandwidth resource allocation based on the semantic deadlines inherent in diffusion-based image inpainting models to optimize outcome quality. In this SGC framework, the transmitter sends two semantic information which are the semantic mask and textual description. Semantic mask provides positional information for transformations and the textual description specifies the content for modification. The receiver applies each semantic information asynchronously to its ongoing process as soon as they are received. Through simulations and visualizations, we demonstrate the effectiveness of the proposed algorithm, highlighting its potential to enhance resource utilization efficiency and offer a new perspective for operating AI applications over communication networks.

\section*{Acknowledgment}
This work has been supported by the 6GARROW project which has received funding from the Institute for Information \& Communications Technology Promotion (IITP) grant funded by the Korean government (MSIT) (No. RS-2024-00435652) and from the Smart Networks and Services Joint Undertaking (SNS JU) under the European Union’s Horizon Europe research and innovation programme under Grant Agreement No 101192194.

\bibliographystyle{IEEEtran}
\bibliography{main}
\end{document}